\documentclass[a4paper,11pt]{article}
\pdfoutput=1 % if your are submitting a pdflatex (i.e. if you have
\usepackage{jinstpub} % for details on the use of the package, please
\usepackage{lineno}
\title{Liquid argon characterization of the X-ARAPUCA with alpha particles, gamma rays and cosmic muons}

\author[a,1]{H.~V.~Souza,\note{Corresponding author.}}
\author[a,1]{E.~Segreto,}
\author[a]{A.~A.~Machado,}
\author[a]{R.~R.~Sarmento,}
\author[b]{M.~C.~Q.~Bazetto,}
\author[c]{L.~Paulucci,}
\author[d]{F.~Marinho,}
\author[b]{V.~L.~Pimentel,}
\author[a]{F.~L.~Demolin,}
\author[a]{G.~de Souza,}
\author[a]{A.~C.~Fauth,}
\author[e]{and M.~A.~Ayala-Torres}

\affiliation[a]{Instituto de Física Gleb Wataghin, Universidade Estadual de Campinas - Unicamp, \\ Rua Sérgio Buarque de Holanda, No 777, CEP 13083-859 Campinas, SP, Brazil}
\affiliation[b]{Centro de Tecnologia da Informação Renato Archer - CTI,\\ Campinas, 13069-901 SP, Brazil}
\affiliation[c]{Universidade Federal do ABC,\\Av. dos Estados, 5001, Santo André, SP, 09210-170, Brazil}
\affiliation[d]{Universidade Federal de São Carlos, Rodovia Anhanguera,\\ km 174, 13604-900, Araras, SP, Brazil}
\affiliation[e]{Centro de Investigación y Estudios Avanzados del IPN, \\ Apartado Postal 14-740, CdMx 07000, Mexico}

\emailAdd{hvsouza@ifi.unicamp.br, segreto@ifi.unicamp.br}

\abstract{The X-ARAPUCA device is the baseline choice for the photon detection system of the first far detector module of the DUNE experiment. We present the results of the first complete characterization of a small scale X-ARAPUCA prototype, which is a slice of a full DUNE module. Its total detection efficiency in liquid argon was measured with three different ionizing radiations: $\alpha$ particles, $\gamma$'s and muons and resulted to be $\sim$2.2\% when the active silicon photomultipliers were biased at +5.0 V of over voltage, corresponding to a Photon Detection Efficiency around 50\% at room temperature.
This value comfortably satisfies the requirements of the first DUNE far detector module (detection efficiency $>$2.0\%) and allows to achieve an energy resolution comparable to the one achievable with the Time Projection Chambers for energies below 10~MeV, which is the region relevant for Supernova neutrino detection.}


\begin{document}
\maketitle
\flushbottom

\section{Introduction}
\label{sec:intro}
The Deep Underground Neutrino Experiment (DUNE) is a long baseline neutrino experiment which will address, in the next years, some of the most relevant questions in neutrino physics, such as the mass hierarchy and the CP violation in the leptonic sector. The huge mass of the far detector of about 40 kt, in its final configuration,  will also allow to carry out a rich program of non-beam physics such as proton decay searches, supernova and solar neutrino detection. The ambitious DUNE physics program will be enabled by the combination of three key elements: (i) the most intense neutrino beam ever built - located at Fermilab (Batavia, Illinois - USA), (ii) a fine grained near detector - located a few hundred meters downstream of the neutrino source, (iii) a massive far detector installed in the Sanford Underground Research Facility (SURF), 1,300~km distant from the neutrino source and 1,500~m underground. 
The far detector will use the technique of liquid argon time projection chamber (LArTPC) which allows to perform a complete tri-dimensional and calorimetric reconstruction of the charged particle tracks produced by the neutrino interaction through the detection of the scintillation light and of the ionization charge. The scintillation light is important to determine the time, T$_0$, at which the ionization electrons start drifting toward the anode sense wires, under the action of a very uniform electric field of about 500~V/cm. The drift time, between T$_0$ and the arrival of the ionization electrons on the anode wires, is used to reconstruct the coordinate along the drift direction with a resolution better than 1~mm, which is important to correct for the absorption of the ionization charge along the drift and to fiducialize the active volume of the detector. Scintillation light detection will allow to perform precise calorimetric measurements, since at a typical electric field of 500~V/cm, about half of the energy deposited in LAr is converted into scintillation photons. Dedicate Monte Carlo studies have shown that with the level of photon detection efficiency foreseen for the first DUNE far detector module, it will be possible to achieve an energy resolution with light, which is at least comparable with the one achievable with charge at energies below few tens of~MeV - the region relevant for supernova and solar neutrinos.

The entire DUNE far detector consists of four modules of 10 kt fiducial mass. The Photon Detection System (PDS) of the first DUNE far detector module will be based on the X-ARAPUCA technology, which is an evolution and improvement of the ARAPUCAs~\cite{propostaARA}, extensively tested in protoDUNE during its first RUN at CERN~\cite{first_results_protodune}. The PDS will be composed by 1,500 independent bar-shaped modules, with approximate dimensions of 210~cm~$\times$~12~cm and a thickness of 23~mm which allows them to be installed between the anode wire planes~\cite{DUNE_vol4}.

\begin{figure}[htbp]
    \centering
     \includegraphics[width=6.0cm]{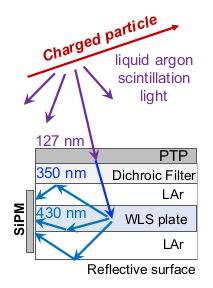}
    \caption{\label{fig:X-ARAPUCA} Pictorial representation of the X-ARAPUCA working principle. LAr scintillation photons at 127~nm are absorbed by a thin para-Terphenyl layer and re-emitted at 350~nm. They enter the reflective cavity through the dichroic filter (cut-off at 400~nm) and are wave-shifted by the WLS plate at a wavelength around 430~nm. A fraction of photons is trapped through total internal reflection inside the WLS plate and the remaining fraction through the standard ARAPUCA mechanism. The active light sensors (SiPM) are installed on the lateral sides of the reflective cavity.}
\end{figure}

The X-ARAPUCA is a light trap and its working principle is schematically illustrated in Figure~\ref{fig:X-ARAPUCA}. LAr scintillation light is absorbed by the thin layer of para-Terphenyl (PTP)~\cite{pTP} which is deposited by vacuum evaporation on the external side of a dichroic filter, which represents the acceptance window of the device. PTP down-converts the absorbed radiation at 127~nm to around 350~nm, with high efficiency. The cut-off wavelength of the dichroic filter is chosen to be 400~nm, in order to allow the PTP shifted light to get inside the reflective X-ARAPUCA cavity, which hosts a thin wavelength-shifting plate (WLS plate) with same surface area of the filter. The light emitted by PTP is absorbed by the WLS plate and re-emitted at 430~nm, in the spectral region where the filter is highly reflective. An array of active silicon photon-multipliers (SiPM) are installed on the sides of the reflective cavity. A fraction of the light re-emitted by the WLS plate is guided towards the SiPM array by total internal reflection inside the plate, since the refractive index of LAr, which fills the cavity, at the emission wavelength of the WLS plate is about 1.2, while that of the WLS plate is 1.58~\cite{eljen_286}. The fraction of photons which does not undergo total internal reflection is trapped through the standard ARAPUCA mechanism. The internal surfaces of the cavity and the lateral edges of the WLS plate are lined with highly reflective dielectric foils.

In this work we describe the first X-ARAPUCA LAr tests, which happened at the Laborat\'orio de Leptons of UNICAMP from October 2019 to March 2020. The X-ARAPUCA prototype was exposed to well known $\alpha$ and $\gamma$ sources and to cosmic muons in order to precisely estimate its absolute detection efficiency. These tests represent a fundamental step in the validation process of the X-ARAPUCA technology for the DUNE first far detector module.

\section{Experimental set-up}
The X-ARAPUCA prototype under test has external dimensions of 9.6~cm$\times$12.5~cm and hosts one single dichroic filter of area 8~cm$\times$10~cm. The filter is designed to have a cut-off wavelength of 400~nm, a transmittance band which goes from 300~nm to 400~nm (transmissivity $\simeq$~90~\%) and a reflective band ranging from 400~nm to 500~nm (reflectivity $\simeq$~98~\%).
The filter is coated, by vacuum evaporation, with a layer of PTP with a surface thickness $\simeq$ 400~$\mu$g~cm$^{-2}$. PTP absorbs 127~nm scintillation light and re-emits it around 350~nm~\cite{pTP}, well inside the transmission band of the filter. The inner cavity is 6~mm deep and hosts the WLS plate (Eljen EJ-286~\cite{eljen_286}), which has a thickness of 3.5~mm. The absorption spectrum of the WLS plate is well matched with the emission spectrum of the PTP with maximum absorbance around 355~nm, while its emission is centered around 430~nm, inside the reflective band of the filter. Two arrays of 4 SiPM each are installed on the 10~cm long lateral sides of the cavity. The SiPMs are Hamamatsu S13360-6050VE with active area of 6$\times6$~mm$^2$~\cite{hmmt_s13360}. The internal surfaces of the cavity are lined with 3M ESR VIKUITI reflective foils~\cite{vikuiti}, as well as the lateral sides of the WLS plate, with the exception of the regions where the SiPMs are installed.   

%EDITAR!!! integrar
%The WLS light guide used in this module is the EJ286  produced by ELJEN\footnote{Eljen Co. \url{https://eljentechnology.com/}.}, a blue-emitting WLS plastic with maximum absorbance around 355~nm and remitting around 430~nm. The SiPMs produced by Hamamatsu (S13360-6050VE)~\cite{hmmt_s13360} are the active photo-sensitive elements used to detect the converted and trapped photons.

\subsection{X-ARAPUCA components preparation}
%The X-ARAPUCA module is composed by a mechanical frame (made by TVE G10 (CITE)), a short pass dichroic filter (400~nm cutoff) coated with p-Terphenyl (pTP), a wavelength shifting (WLS) light guide with a polyvinyl toluene (PVT) matrix and as an active photo-sensitive element the silicon photomultipliers (SiPMs).
All the components of the X-ARAPUCA prototype under test have been produced following the DUNE Collaboration criteria and prepared under a rigorous quality control before the assembly.

\begin{figure}[htbp]
    \centering
     \includegraphics[width=0.7\textwidth]{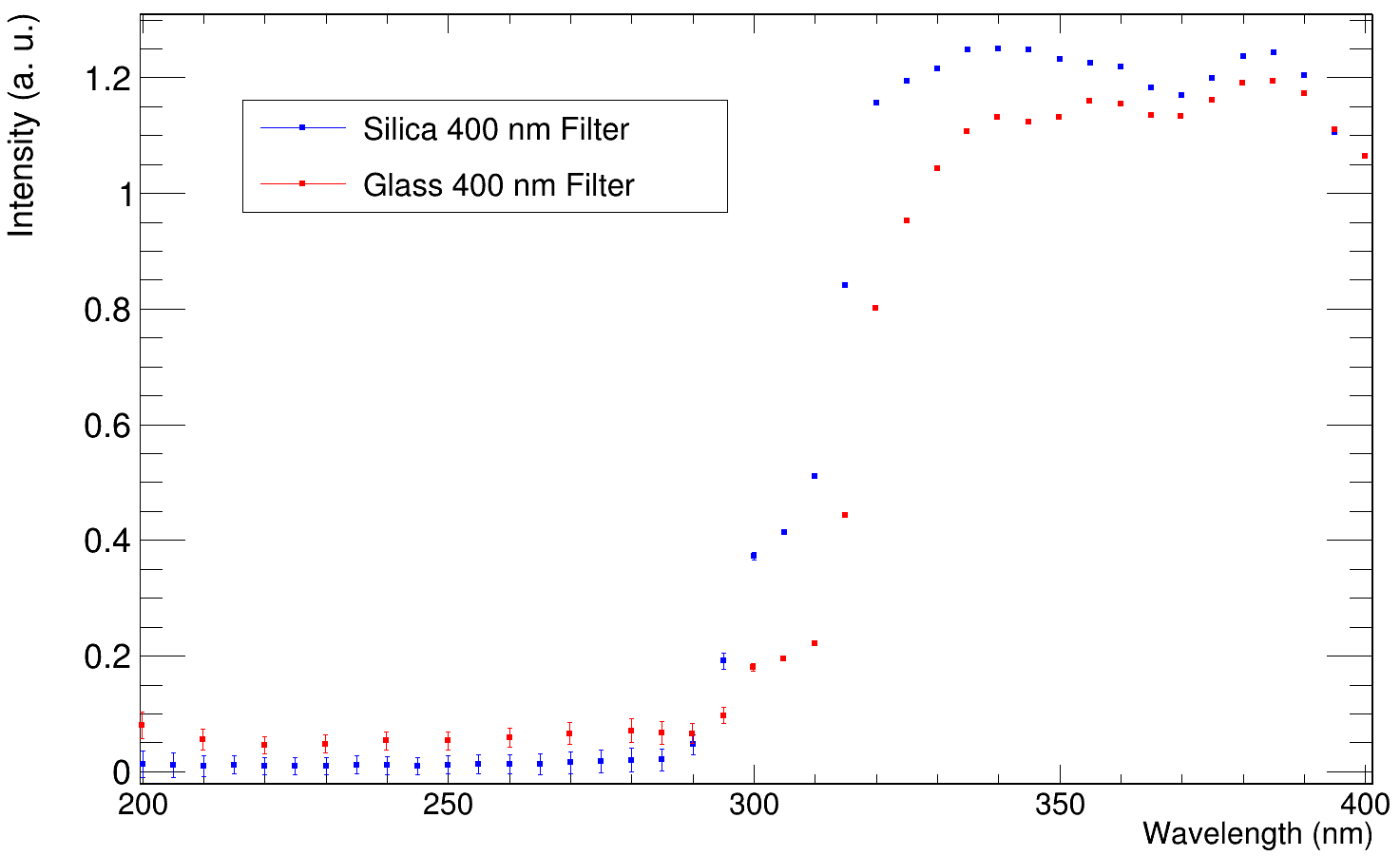}
    \caption{\label{fig:silica_and_glass} Transmittance spectra of fused silica and optical glass (B270).}
\end{figure}

The frames are produced by the Brazilian company MATOOL\footnote{Matool, \url{https://www.matool.com.br/}.} and were cleaned in an ultrasonic bath, dried in a vacuum oven and stored in a clean room. The dichroic filters have been produced by the Brazilian company OPTO\footnote{Opto, \url{https://www.opto.com.br/}.}. They are built as  multi-layer interference filters on top of an appropriate substrate, which needs to be highly transparent to wavelength range corresponding to the emission spectrum of PTP (between 350~nm and 400~nm). Two substrates have been investigated: fused silica and optical glass (B270) and both  were tested in a vacuum monochromator. The two transmittance spectra are shown in Figure~\ref{fig:silica_and_glass}.
The two substrates have comparable performances in the range from 350~nm to 400~nm. The optical glass was chosen because it is easier to cut, the PTP adherence on it is better and the robustness of the film at LN$_2$ temperature is higher than fused silica. Furthermore optical glass is sensibly cheaper.

These dichroic filters were cleaned at CTI Renato Archer\footnote{Centro de Tecnologia Renato Archer, \url{http://www.cti.gov.br}.} using a well established protocol (developed in close collaboration with the thin film laboratory at Fermilab - Batavia, USA) that represents an important step to reach the optimal adherence of the PTP film during the evaporation process and to ensure the robustness of the film during the cool down to LAr temperature. 

After the cleaning procedure, the filters were coated with PTP in a vacuum chamber. 
The evaporation process is performed in high vacuum, with a residual pressure inside the chamber at the level of 10$^{-6}$~mbar. The PTP powder is heated inside a copper Knudsen crucible at a temperature of 220$^o$~C and is vaporized from a hole with a diameter of $\simeq$1~mm. The thickness of the film is monitored with a quartz oscillator installed inside the chamber, close to the filters.

%The PTP shifts the Vacuum Ultra-Violet (VUV) photons wavelength $\sim$127~nm to around 350~nm, in this range the filter becomes transparent to these photons. After the entrance of these photons in the X-ARAPUCA, they will be shifted again for the WLS light guide to a wavelength above the cutoff of the filter, this process let the photon trapped until be detected by SiPM~\cite{x_arapuca}.
%(ARRUMAR!)

\subsection{Cryogenic setup and data acquisition}
\label{sec:cryo_daq}

\begin{figure}[htbp]
    \centering
     \includegraphics[width=0.99\textwidth]{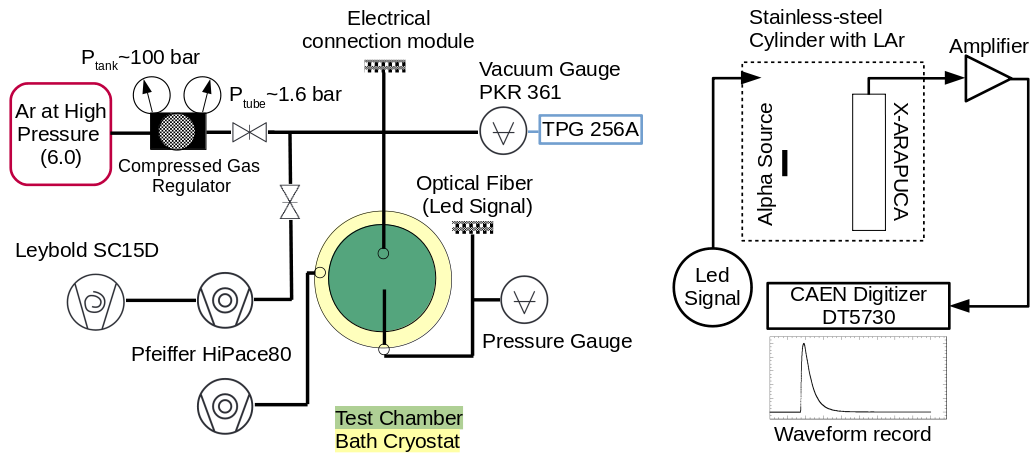}
    \caption{\label{fig:schmatics} (Left) Schematic diagram of the cryogenic setup. (Right) Schematic diagram of the DAQ.}
\end{figure}

\begin{figure}[htbp]
\centering
\includegraphics[width=.35\textwidth]{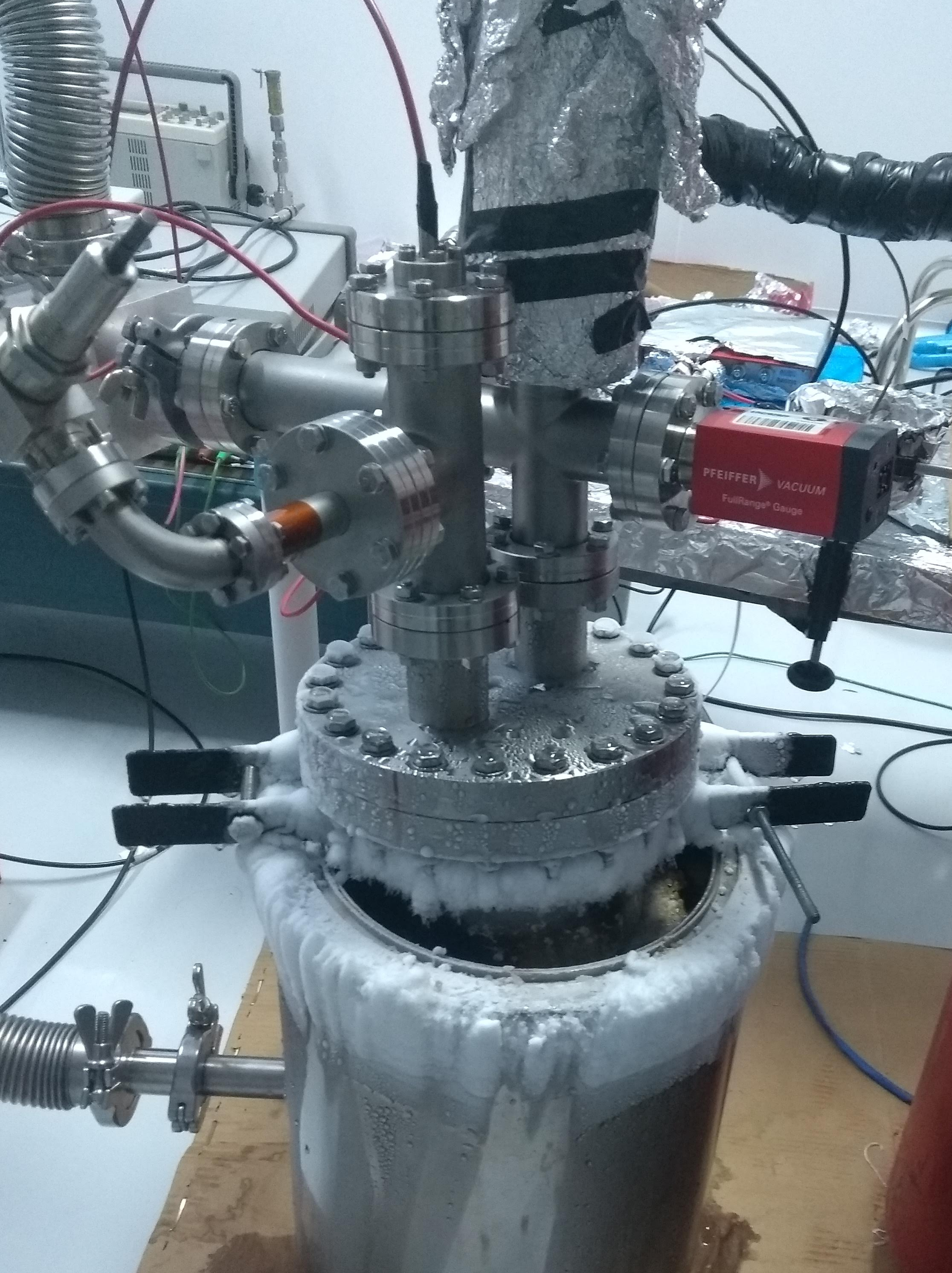}
\quad
\includegraphics[width=.35\textwidth]{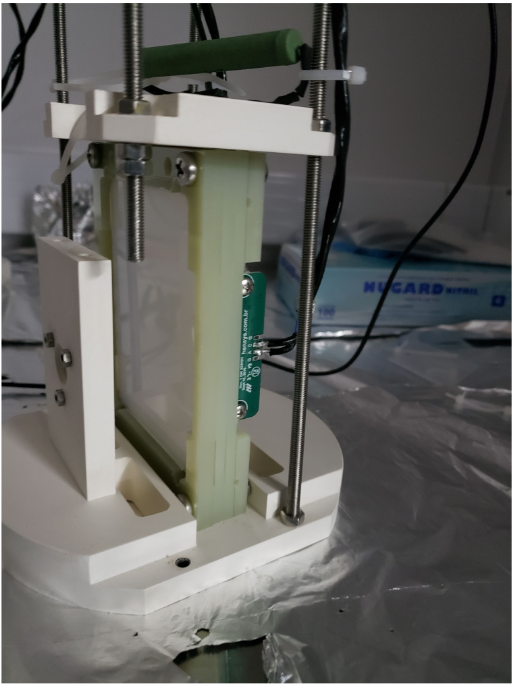}
\caption{\label{fig:setup_photos} (Left) Picture of the cryogenic setup during $\alpha$-source run. (Right) the prototype when
installed in its sustaining mechanical structure together with the $\alpha$-source holder. Level sensor can be seen above the prototype in green.}
\end{figure}

A dedicated cryogenic set-up was assembled to carry on the test of the efficiency of this X-ARAPUCA prototype. Figure~\ref{fig:schmatics} (Left) shows a schematic diagram of the set-up. The inner stainless-steel cylinder (green) is pumped down to a pressure $\sim$$10^{-4}$~mbar at room temperature. Then, a cryogenic pump down, to a pressure of $\sim$$10^{-6}$~mbar, is performed by filling the external open thermal bath (yellow) with liquid argon (LAr). The pump's gate valve is closed and Gas Argon~6.0\footnote{Gas Argon 6.0 refers to a contaminants level equal or below 1~ppm.} is injected in the cylinder, keeping the inner pressure $1<P_{\text{work}}<1.4$~bar. 
%The gas argon {\bf Suggestion: Argon gas }
Argon gas (GAr) is liquefied thanks to the heat exchange with the thermal bath and a level sensor installed inside the stainless-steel cylinder ensures that the 
%{\bf Suggestion: X-ARAPUCA is completely submerged in LAr} LAr completely submerges the X-ARAPUCA
X-ARAPUCA is completely submerged in LAr (see Figure~\ref{fig:setup_photos}). A vacuum tight optical feed-through was installed inside the stainless-steel cylinder to generate pulsed LED flashes in order to properly calibrate the system (see section~\ref{sec:calibration})

%In order to properly calibrate the system, a vacuum tight optical feed-through was installed to bring the flashes of a pulsed LED inside the stainless-steel cylinder (see section~\ref{sec:calibration}) {\bf Suggestion: A vacuum tight optical feed-through was installed inside the stainless-steel cylinder to generate pulsed LED flashes in order to properly calibrate the system (see section~\ref{sec:calibration})}. %The procedure is done to lower the N$_2$ and O$_2$ contamination~\cite{nitrogen_contamination_roberto}.

%Inside the stainless-steal cylinder, 
The X-ARAPUCA is installed facing an $\alpha$-source 3~cm away from the center of the dichroic filter, as shown in the diagram of Figure~\ref{fig:schmatics} (Right) and the picture of Figure~\ref{fig:setup_photos} (Right). A Polyvinyl chloride (PVC) support was built to hold %vertically the X-ARAPUCA {\bf Suggestion: the X-ARAPUCA in a vertical position}
the X-ARAPUCA in a vertical position and to host the $\alpha$ source. For the $\gamma$-rays and cosmic-$\mu$ tests, the $\alpha$-source was removed but the holder was kept in place to maintain the same geometrical configuration and not to bias the measurements with different amount of environmental reflected light. The light was read-out by two arrays of SiPMs. The signal is amplified with a gain of 32~dB by a custom, low-noise, two channels, pre-amplifier produced  by the Brazilian Company Hensys and read-out with a CAEN DT5730 digitizer with sampling frequency of 500~MHz.


The $\alpha$-source is made of an alloy of aluminium and natural uranium in the form of a thin disk with thickness of 0.14~mm and diameter of 1~cm. The energy, relative abundance and parent nucleus of the $\alpha$ source are shown in Table~\ref{tab:alpha_spectrum}. For the $\gamma$-rays test, the $\alpha$-source was removed and a $^{60}$Co source was placed outside the cryostat in a position where the emission of the source faces the X-ARAPUCA acceptance window. %This choice is arbitrary and it was made to allow a better simulation and to try to maximize the count rate.
\begin{table}[htbp]
	\centering
	\caption{\label{tab:alpha_spectrum}Energy, relative intensity and parent nucleus of the $\alpha$ particles emitted by the natural uranium~\cite{alpha_spectrum_tab}.}
	\smallskip
	\begin{tabular}{|c|c|c|}
		\hline
		$\alpha$ energy (MeV) & relative intensity & parent nucleus               \\ \hline
		4.187                 & 48.9\%             & $^{238}$U \\
		4.464                 & 2.2\%              & $^{235}$U \\
		4.759                 & 48.9\%             & $^{234}$U \\ \hline
	\end{tabular}
\end{table}

The data were taken during three different tests for each ionizing radiation. The ADC read-out was done whenever one of the two arrays exceeded a fixed threshold, which corresponded approximately to 5~photo-electrons. However, a fine tuning of the threshold was done manually in order to achieve  a good balance between trigger rate and spectrum resolution. Triggered events were recorded for 18~$\mu$s with 3.6~$\mu$s of pretrigger. 

Dedicated calibration runs were performed for each one of the bias voltage which were used for the SiPMs. A LED was flashed using pulses of approximately 50~ns, frequency of 1~kHz and amplitude set in order to produce signals of few photo-electrons. The acquisition of the signals was triggered with the external trigger of the Pulse Generator \textit{Agilent} with the same record length of 18~$\mu$s used for source signals.

Each data sample contains about 40,000 events both for the radioactive
sources and for the calibration. The only exceptions were the runs with the $^{60}$Co $\gamma$ source, where data were taken during a fixed period of time of about 30 minutes. Data were taken with and without the $^{60}$Co source, since its activity was quite low and it was necessary to subtract the background in order to extract a clean Compton spectrum. 
%in two forms, with and without the $^{60}$Co source. This is performed so the noise and background can be properly subtracted from the spectrum as describe in section~\ref{sec:analysis}.

\section{Data analysis and results}

\subsection{SiPM calibration}
\subsubsection{Single photon response}
\label{sec:calibration}
The SiPMs were operated in reversed bias with a few volts over the breakdown voltage of 43.0~$\pm$~0.2~V. Among the different overvoltages (O.V.) used, the values of +5.0 and +5.5~V~($\pm 0.2$~V)~O.V.\ were selected due to the quality of the calibration runs and of the spectra collected.
%EDIT!!! Melhorar! O.V.

\begin{figure}[ht]
\centering
\includegraphics[width=.45\textwidth]{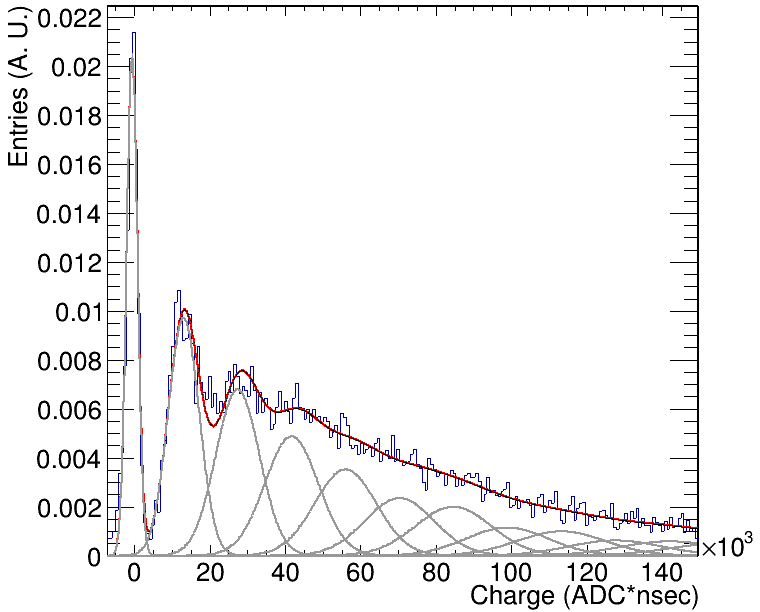}
\includegraphics[width=.44\textwidth]{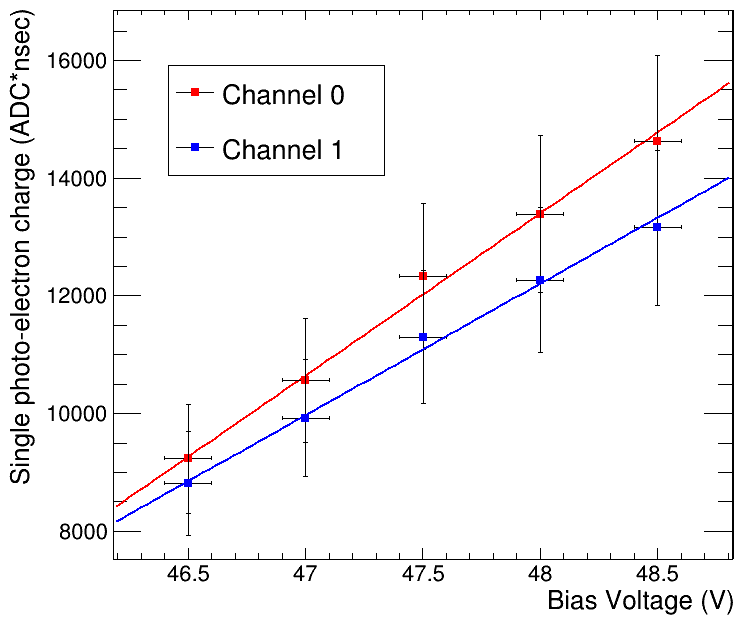}
\caption{\label{fig:sphe} (Left) Single photo-electron response obtained during calibration. The distance between the first and second peaks after the noise 
%The distance between the first peak after the noise and the second one 
%{\bf Suggestion: The distance between the first and second peaks after the noise or The distance between the first peaks after the noise and the following one }
gives the gain of the SiPM. (Right) Gain versus bias voltage for the two channels of the X-ARAPUCA. Red and blue lines represent linear fits of the experimental points. The breakdown voltage was found to be 43.0~$\pm$~0.2~V.}
\end{figure}

%In order to properly evaluate the amount of photo-electrons detected by the device, the calibration data were used to retrieve the single photon-electron (SPHE) response. Each waveform acquired with the LED was integrated over 900~ns starting from the pre-trigger point (3.6~$\mu$s)~\cite{SiPM_better} EXPLICAR MELHOR.
In order to properly evaluate the amount of photo-electrons detected by the device, the calibration data were used to retrieve the single photon-electron (SPHE) response. The waveforms, acquired with the LED as external trigger, are integrated for 900~ns (decay time of SPHE signal $\sim$300~ns) starting from the trigger point (3.6~$\mu$s)~\cite{SiPM_better}. 
%so the charge of the photo-electrons response are taken.
A spectrum obtained with a bias voltage of 48.0~V (+5.0~$\pm$~0.2 O.V.) is shown in Figure~\ref{fig:sphe}. The fit is performed with $N+1$ Gaussian distributions, where the first one corresponds to the electronic noise and the others correspond to first, second and $N$-th photo-electron responses. The noise and SPHE Gaussian distributions have 3 free parameters each (mean value, standard deviation and normalization constant). The mean values and standard deviations of the other peaks are constrained by the following conditions:
\begin{itemize}
    \item the mean value of the {\it N-th} peak is given by $N~\times~x_1$, where $x_1$ is the mean value of the single photo-electron peak;
    \item the standard deviation of the {\it N-th} peak is given by $\sqrt{N}~\times~\sigma_1$, where $\sigma_1$ is the standard deviation of the single photo-electron peak;
\end{itemize}
%all the others are  constrained by the conditions:
%of having the standard deviation equal to the square root of $N$ times the standard deviation of the first SPHE peak. The second photo-electron mean value is fitted and all others peaks will have the mean value fixed at $N$ times the distance between the one photo-electron and two photo-electrons.
The gain was considered to be the distance between the first and second photo-electron. The result for different bias voltages is shown in Figure~\ref{fig:sphe} for the two channels of the X-ARAPUCA. A linear fit is shown together with the data. 
%The gain of the SiPM, together with the cross-talk and after-pulse, consist the main source of uncertainties. 

\subsubsection{Cross-talk probability estimation}\label{crosstalk}

An estimation of the cross-talk probability and of the relative correction factor was performed using the method described in~\cite{Cross_talk_vino}. The probability of detecting $k$ photo-electron is given by the total number of events which contains $k$ photo-electrons, divided by the total number of events. The integration of the $k$-th Gaussian distribution from Figure~\ref{fig:sphe} gives the number of events with $k$ photo-electrons. The probability distribution, $P_k$, of detecting $k$ photo-electrons is described by the distribution:
\begin{equation}
\label{eq:ct_vino}
\begin{split}
P_k(L,p) &= exp(-L)\sum_{i=0}^{k} B_{i,k} \cdot [L(1-p)]^i \cdot p^{k-i}
\\
\text{where,}
\\
B_{i,k} &= \left\{
  \begin{array}{lr}
    \qquad 1 \quad \qquad \text{if } i = 0 \text{ and } k = 0 \\
    \qquad 0 \quad \qquad \text{if } i = 0 \text{ and } k > 0 \\
    \frac{(k-1)!}{i!(i-1)!(k-i)!} \quad \text{otherwise}
  \end{array}
\right.
\end{split}
\end{equation}
where $L$ is the mean value of the Poisson distribution obtained in the case of $p=0$ and $p$ is the cross-talk probability~\cite{Cross_talk_vino}.

\begin{figure}[htbp]
    \centering
    \includegraphics[width=0.45\textwidth]{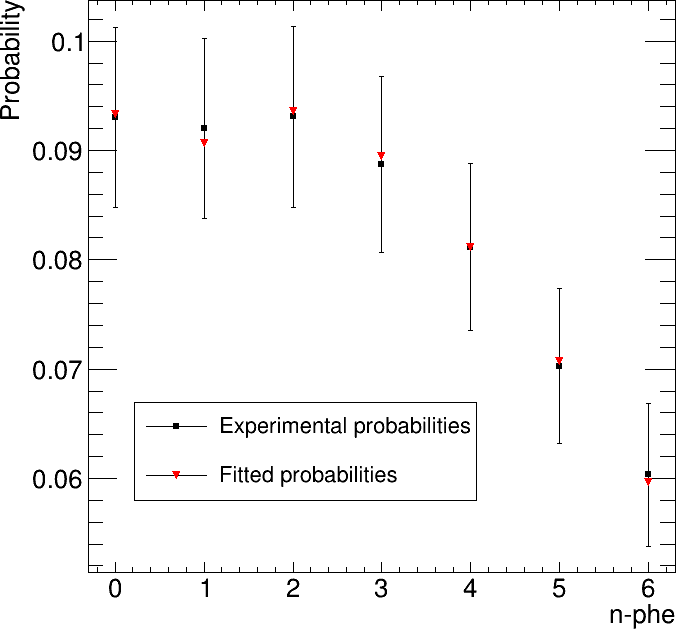}
    \includegraphics[width=0.45\textwidth]{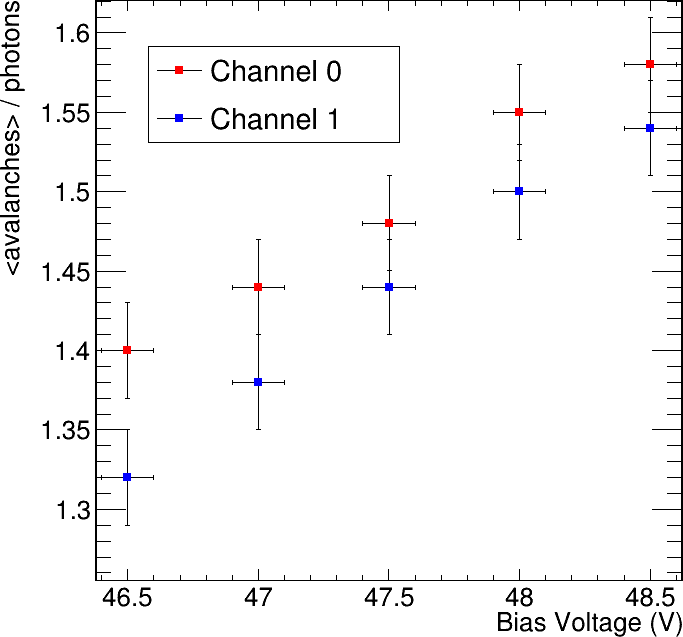}
    \caption{\label{fig:CT_vinos}(Left) Probability of detecting $n$ photo-electrons (black squares). A $\chi^2$ minimization of the experimental points with equation~\ref{eq:ct_vino}~\cite{Cross_talk_vino} was performed and the result is shown as red triangles. (Right) Avalanches per photons versus bias voltage for each channel.}
\end{figure}

%A $\chi^2$ minimization was performed between data and the probability from equation~\ref{eq:ct_vino}, were $L$ and $p$ were taken as parameters and a scaling factor was introduced. 
Normalized calibration spectra were fitted with equation~\ref{eq:ct_vino} via a $\chi^2$ minimization, where $L$ and $p$ were considered as free parameters.
One of the results from the fitting procedures is shown in Figure~\ref{fig:CT_vinos} (Left).
The plot of the number of SiPM avalanches per photon ($1+p$) vs.\ the bias voltage is shown in Figure~\ref{fig:CT_vinos} (Right) for each channel. These values are used as correction factors when we convert the number of detected avalanches into photo-electrons. It ranges between 1.32 and 1.58 avalanches per photons and is consistent with what has been found in the first run of the  protoDUNE detector~\cite{first_results_protodune}.
%The results are compatible considering the uncertainties for all the three different ionizing radiations ($\alpha$-particles, $\gamma$-rays and $\mu$), which makes the X-ARAPUCA efficiency estimation solid. In fact, it does not depend on the particle type nor on the topology of energy deposition, which is distinct for the three sources.

\subsection{Experimental Setup - Monte Carlo Simulation} 
\label{sec:mc}
In order to properly evaluate the efficiency of the X-ARAPUCA prototype, a dedicated GEANT4 Monte Carlo simulation was developed for each type of ionizing radiation considered. All the dimensions of the experimental setup were carefully measured and implemented: the stainless-steal cylinder and cryostat, the X-ARAPUCA and $\alpha$-source PVC support (Figure~\ref{fig:setup_photos}) and the X-ARAPUCA prototype.

The liquid argon scintillation light yield was set at 51,000 photons per~MeV ($\gamma/$MeV)~\cite{LAr_fund_properties,abudance_dependence} times the quenching factor, which was set at 0.71 for the $\alpha$-particles and at 0.78 for electrons and muons~\cite{LAr_fund_properties,model_nuclear_recoil_nl,x_arapuca}. 

\begin{figure}[htbp]
    \centering
    \includegraphics[width=0.65\textwidth]{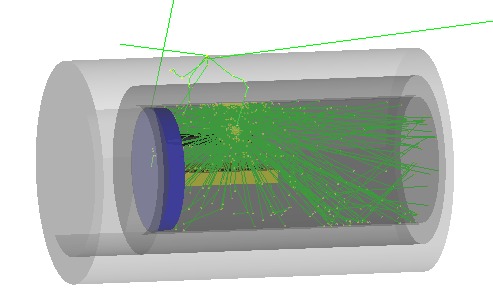}
    \caption{\label{fig:ex_simu} Example of the experimental setup with simulated gamma source outside the cryostat and the optical photons (green lines) reaching the X-ARAPUCA.}
\end{figure}

Alpha particles were emitted uniformly and isotropically inside the aluminium disc with the energy distribution given in Table~\ref{tab:alpha_spectrum}. The $\alpha$-particles lose a fraction of their initial energy inside the aluminum disc (which goes undetected) and leave the aluminium alloy with a continuous residual energy distribution~\cite{alpha_equation}. For $\gamma$-rays, the $^{60}$Co source was positioned outside the cryostat and the emission was chosen to be  isotropic and with two emission lines, one at 1.17~MeV and the other at 1.33~MeV~\cite{cobalt_60}. One example is shown in Figure~\ref{fig:ex_simu}, with the gamma source placed outside the cryostat and optical photons (green) generated in the LAr reaching the X-ARAPUCA (yellow).
Cosmic muons were generated uniformly in a 4 meters diameter disc positioned 50~cm above the cryostat, with a $\cos^2{\theta}$ angular distribution (where $\theta$ is the zenith angle), with an uniform azimuth angle distribution and a fixed energy of 4~GeV, which corresponds to the average muons' energy at ground level~\cite{pdg,MuonsCecchini2012}. The reflectivity of the internal (electro-polished) stainless-steel surfaces  was set at 20\% for the scintillation photons. The final Monte Carlo output is the number of photons that reaches the X-ARAPUCA acceptance window per particle generated. 

\subsection{Efficiency analysis}
\label{sec:analysis}

\begin{figure}[htbp]
\centering 
\includegraphics[width=.48\textwidth]{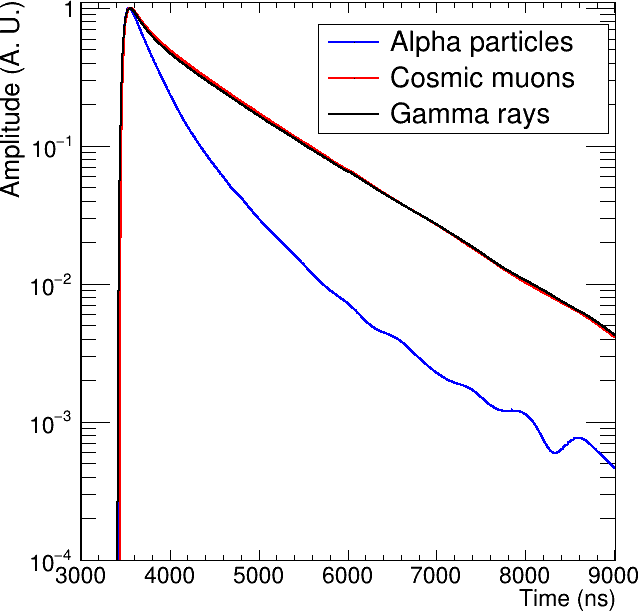}
\includegraphics[width=.48\textwidth]{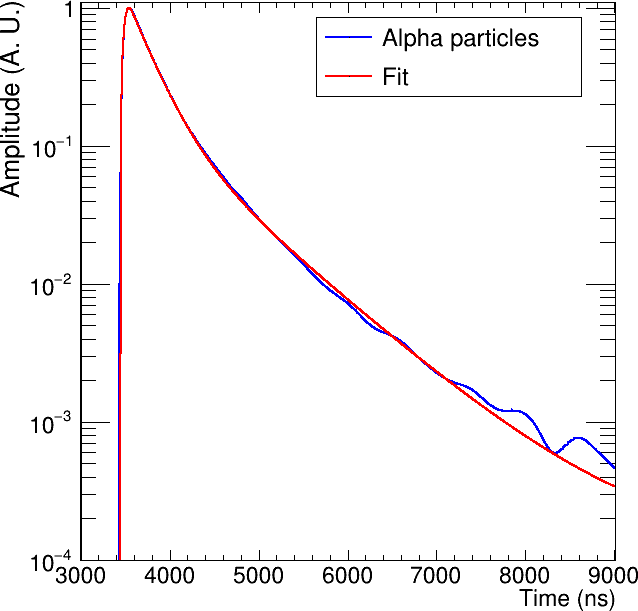}
\caption{\label{fig:mean_signals} (Left) Average signal for the three different particles. (Right) Average signal of $\alpha$-particles fitted with equation~\ref{eq:signal_fit}}.
\end{figure}

An analysis on the average scintillation signal for each run was performed in order to evaluate the LAr purity, validate the tests and correct the total amount of  photo-electrons detected. Figure~\ref{fig:mean_signals} (Left) shows the average signal for $\alpha$-particles (red), $\gamma$-rays (black) and cosmic muons (blue). 
%$\alpha$ particles and $\gamma$/muons show pretty different values of the  ratio of the fast and slow scintillation components, as expected ~\cite{LAr_fund_properties,ICARUS,abudance_dependence}. 
It has been shown  that N$_2$ and O$_2$ contamination, at the level a few ppm, can significantly quench the slow scintillation emission, thus reducing the total photon yield~\cite{nitrogen_contamination_roberto}. 
In order to correct for this effect, an overall fit of the average scintillation waveforms is performed and the value of the slow scintillation decay time is extracted. This value is used to extrapolate the number of photo-electrons which would have been detected in the case of perfectly pure LAr.
%Once the slow component plays a major for the amount of light in muons and electrons scintillation, nitrogen contamination significantly decreases the amount of light detected. To verify if the purity achieved with the cleaning process and GAr 6.0 was properly, we analyzed the slow time component for alphas, muons and gammas. 

Equation~\ref{eq:signal_fit} is used to fit the waveform as presented in Figure~\ref{fig:mean_signals} (Right), where $\tau_f$ and $\tau_s$ are the fast and slow components and $A_f + A_s = 1$ are the relative amplitudes~\cite{Segreto_2021}. The third term of the sum is a component related to a possible PTP delayed emission~\cite{Ettore_tpb}, with $N$ and $A$ as constants depending on the nature of the scintillator and $t_a$ is the relaxation time. $A$ and $t_a$ where constrained to vary between $0.22 - 0.55$ and $20 - 100$~ns, respectively. The fit is performed with a $\chi^2$ minimization of the signal with the light pulses of eq.~\ref{eq:signal_fit}, convoluted with a Gaussian, which takes into account the effect of the read-out electronics.
\begin{equation}
\label{eq:signal_fit}
L(t) = \frac{A_f}{\tau_f} e^{-t/\tau_f} + \frac{A_s}{\tau_s} e^{-t/\tau_s} + \frac{N}{[1+A\;ln(1+t/t_a)]^2 (1+t/t_a)}
\end{equation}

%The values for $\alpha$-particles were found to be $A_s = 0.66 \pm 0.01$  and $\tau_s = 780 \pm 10$~ns. For $\gamma$-rays we found $A_s = 0.11 \pm 0.1$ and $\tau_s = 1100 \pm 10$~ns and for muons we found $A_s = 0.12 \pm 0.1$ and $\tau_s = 1050 \pm 10$~ns. The fast component is not possible to be retrieved because of the SiPM time response, so values range from 210 to 250~ns.

%Comparing the values of the slow decay times, found with this fitting procedure, {\bf Suggestion: Comparing the slow decay times obtained with this fitting procedure}
Comparing the slow decay times obtained with this fitting procedure with the expected values~\cite{nitrogen_contamination_roberto}, we determined the correction factors of 1.16, 1.43 and 1.47 for $\alpha$-particles, $\gamma$-rays and cosmic muons, respectively, to be applied to the total amount of detected photo-electrons. 
%%% THIS IS NOT CLEAR TO ME ---- REMOVED FOR THE MOMENT
%One should point out that although we found a value of $A_s \sim $ 0.1 for $\mu$ and $\gamma$-rays, $A_s$ were assumed as 0.77 and $A_f = 23$ for $\mu$ and $\gamma$-rays. For the $\alpha$-particles, the values where assumed as $A_s = 0.23$ and $A_f = 0.77$, with a lower correction~\cite{nitrogen_contamination_roberto,abudance_dependence}.

The efficiency of the X-ARAPUCA has been determined through a fitting procedure of the Monte Carlo spectrum of the number of photons impinging on the acceptance window to the experimental spectrum of the number of detected photo-electrons 
%{\bf Suggestion: Here it seems should be the opposite, the ratio between detected over landing ph. Maybe rewrite it as The efficiency of the X-ARAPUCA has been determined through a fitting procedure of the experimental spectrum of the number of detected photo-electron and the Monte Carlo spectrum of the number of photons impinging on the acceptance window, as the ratio between these two quantities.}
%Resposta: here it is not the ratio, so I changed "over the" to "to the" 
(the TMinuit package from Root was used~\cite{root_cern}).
%The comparison between experimental data and Monte Carlo simulation was %done through a $\chi^2$ minimization of the detected photo-electron spectra 
%The fitting procedure was done through a $\chi^2$ minimization of the detected photo-electron spectra using the TMinuit package from Root CITE!!!.
The total amount of detected photo-electrons is computed integrating the digitized waveform over 15.3~$\mu$s starting from 300~ns before the onset. The integral value is multiplied by the correction factor due to LAr purity and then divided by the gain of the SiPM times the cross-talk correction factor. 

In the case of $\alpha$-particles, a data selection using $F_{\text{prompt}}$ ratio~\cite{LAr_arapuca_test} was performed in order to discriminate genuine $\alpha$ signals from muon's ones. An integration interval of 500~ns for the fast component, in the calculation of  $F_{\text{promp}}$, was chosen. This is because of the response function of the SiPMs which spreads the fast component over hundreds of~ns.

The $\gamma$-rays spectrum was obtained by subtracting the histogram of the photo-electrons detected with the source to the histogram taken without the source. This is needed to remove events related to cosmic muons and background. Because of this subtraction, there are less entries in the $\gamma$-ray spectra than for the other sources. 

As described in section~\ref{sec:mc}, the output of the GEANT4 simulation is the amount of photons reaching the acceptance window of the detector per each simulated event. The fitting procedure of the Monte Carlo spectrum over the experimental one foresees the use of two free parameters. One is the normalization constant of the Monte Carlo spectrum and the other is the scale factor between photons and photo-electrons, which is directly the efficiency of the X-ARAPUCA prototype.

%In this sense, only two parameters differ between the experimental data histogram and the MC output, an horizontal constant multiplying the amount of photons reaching the window and a vertical constant to normalize the histograms. The X-ARAPUCA efficiency is directly related to the horizontal constant. 

\begin{figure}[htbp]
\centering 
\includegraphics[width=.32\textwidth]{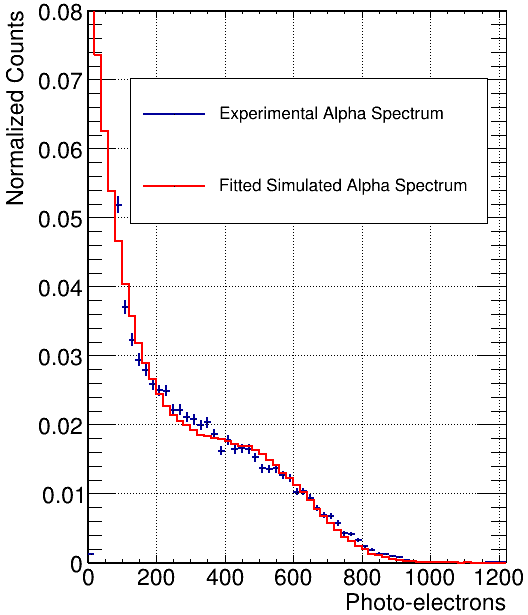}
\includegraphics[width=.31\textwidth]{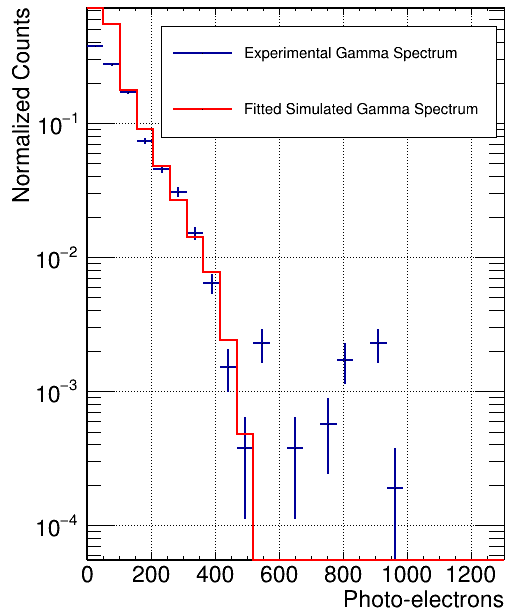}
\includegraphics[width=.31\textwidth]{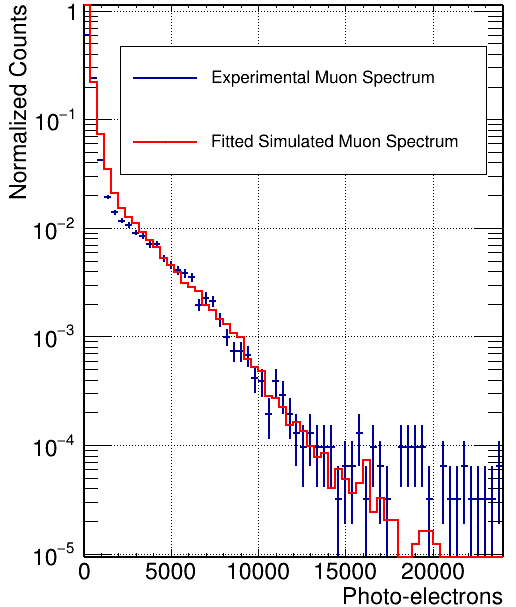}
\caption{\label{fig:fitted_spectrums} Monte Carlo spectrum (red) fitted in the experimental spectrum (blue) for $\alpha$-particles (left), $\gamma$-rays (center) and cosmic muons (right).}
\end{figure}

Figure~\ref{fig:fitted_spectrums} shows the results for the three different ionizing radiations, were the red lines represent the results of the fit of  the MC output on the experimental data (in blue). A Gaussian smearing of the MC $\alpha$ spectrum is performed in order to compensate for the detector energy resolution. An  exponentially decaying function is added to MC spectra to take into account the very low energy background entering the stainless-steel vessel and which is not included in the MC simulation. The results for the X-ARAPUCA efficiency are summarized in table~\ref{tab:results}. The cases where the cross-talk is considered or not are both shown.
%The results look pretty consistent {\bf Suggestion: look pretty consistent is not exactly paper language... Substitute by are compatible within X standard deviations, perhaps?}
The results are compatible considering the uncertainties for all the three different ionizing radiations ($\alpha$-particles, $\gamma$-rays and $\mu$), which makes the X-ARAPUCA efficiency estimation solid. In fact, it does not depend on the particle type nor on the topology of energy deposition, which is distinct for the three sources.
%In fact, it results not to depend not only on the particle type but also on the topology of energy deposition which is pretty different for the three sources. {\bf Suggestion: In fact, it does not depend on the particle type nor on the topology of energy deposition, which is distinct for the three sources.}

The two main sources of uncertainties are the single photo-electron calibration and the cross-talk probability. It has been noticed that the S13360-6050VE MPPCs present some after-pulses that degrades the quality of the single photon spectrum, because of this an uncertainty of 15\% due to the SPHE calibration was considered, based on the quality of the fits. The cross-talk calibration had a maximum variation of $\sim$10\% which was conservatively chosen. The combined uncertainty was taken as 20\%.
%EDITAR!!
%We believe that the efficiency at 48.5 V for $\gamma$-rays was found to be lower because of the low statistics of the spectrum. Even with such a variation, we  safely retrieved an efficiency of (2.3 $\rightarrow$ 3.1) $\pm$ 0.5 \% for the X-ARAPUCA prototype.
%NAO ESTA CLARO. EU TRIARIA, MAS SE CONSEGUIMOS EXPLICAR MELHOR PODE FICAR
\begin{table}[htbp]
	\centering
	\caption{\label{tab:results}Efficiency obtained not considering and considering the cross-talk correction.}
	\smallskip
	\begin{tabular}{c|c|c|c|cc}
		\cline{2-4}
		& \begin{tabular}[c]{@{}c@{}}$\mu$ eff. (\%)\end{tabular} & \begin{tabular}[c]{@{}c@{}}$\gamma$ eff. (\%)\end{tabular} & \begin{tabular}[c]{@{}c@{}}$\alpha$ eff. (\%)\end{tabular} &  &  \\ \cline{2-4}
		& \multicolumn{3}{c|}{Without Cross-talk}                                                                                                                                           &  &  \\ \cline{1-4}
		\multicolumn{1}{|c|}{48.0 V} & 3.47~$\pm$~0.52                                          & 3.43~$\pm$~0.51                                           & 3.36~$\pm$~0.50                                       &  &  \\ \cline{1-4}
		\multicolumn{1}{|c|}{48.5 V} & 5.00~$\pm$~0.75                                          & 4.20~$\pm$~0.63                                          & 4.71~$\pm$~0.71                                          &  &  \\ \cline{1-4}
		\multicolumn{1}{|c|}{}       & \multicolumn{3}{c|}{With Cross-talk}                                                                                                                                              &  &  \\ \cline{1-4}
		\multicolumn{1}{|c|}{48.0 V} & 2.33~$\pm$~0.47                                           & 2.32~$\pm$~0.46                                           & 2.20~$\pm$~0.44                                                       &  &  \\ \cline{1-4}
		\multicolumn{1}{|c|}{48.5 V} & 3.13~$\pm$~0.63                                           & 2.65~$\pm$~0.53                                           & 3.01~$\pm$~0.60                                          &  &  \\ \cline{1-4}
	\end{tabular}
\end{table}

The detection efficiency measured with the SiPMs biased at +5.0~V of O.V. (which corresponds to a PDE - Photon Detection Efficiency - of about 50\% at room temperature) is compatible with what reported in~\cite{xarapuca_milano_tests} where the SiPMs (Hamamatsu S14160-6050HS in that case) were biased at +2.7~V of O.V., which corresponds to approximately the same PDE of about 50\%. The detection efficiency measured at +5.5~V of O.V. appears to be a bit overestimated, since it is not perfectly compatible with the expected increase in the PDE of the SiPMs. In particular one of the two read-out channels showed an increase in detection efficiency when going from +5.0~to +5.5~V of O.V. not compatible with the trend observed up to +5.0~V of O.V. The reasons for this inconsistency are under investigation.      

The results presented here come from the last three tests performed in liquid argon. Other tests with $\alpha$-particles were performed and are in agreement with the efficiency presented here~\cite{first_lar_test}. Due to several thermal stresses along these tests, micro-cracks formation has been noticed in the internal WLS slab. A decrease in efficiency has been noticed between the first and last test and was attributed to the micro-cracks in the slab. We believe that two factors can improve the performance of the  X-ARAPUCA and its efficiency: the annealing  of the WLS slab to suppress the formation of micro cracks of the light-guide and a better optical contact between the SiPMs modules and the slab, avoiding the loss of trapped photons.    

\section{X-ARAPUCA Simulation}
\label{sec:mc_arapuca}
\begin{figure}
\begin{center}
\includegraphics[width=0.4\textwidth]{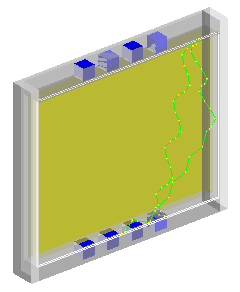}
\caption{X-ARAPUCA view in GEANT4 simulation. The box is shown in gray, SiPMs in blue, and WLS bar in yellow. A simulated photon trajectory is shown in green.}
\label{simulation}
\end{center}
\end{figure}

\begin{figure}
\begin{center}
\includegraphics[width=0.5\textwidth]{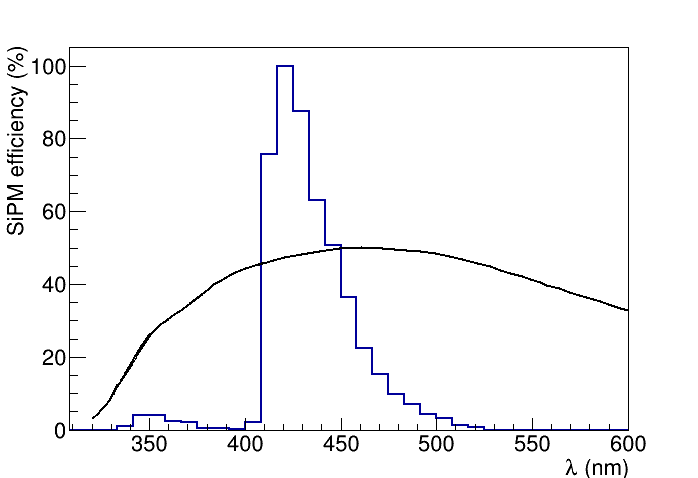}
\caption{Spectrum of photons reaching the SiPMs from the simulation (in blue, from 50000 photons reaching the outside of the X-ARAPUCA window) compared to the detection efficiency of the Hamamatsu S13360-6050VE (in black) taken for 25${^o}$C temperature and a +5.0~V overvoltage.}
\label{SiPMEff}
\end{center}
\end{figure}

A full simulation of the device was performed using the GEANT4 framework~\cite{Agostinelli2003, Allison2016}. The following elements were included in the simulation, and can be seen on fig.~\ref{simulation}:

\begin{itemize}
\item {\bf X-ARAPUCA frame}: A 95.7~mm x 77.7~mm x 6~mm box lined with 3M ESR specular reflector on the inside containing 8 cuts for placing the SiPMs, being 4 on each of its larger sides. Reflectivity to photons in the visible region of the spectrum is taken to be 98\%.
\item {\bf Acceptance window:} Composed of a dichroic filter on an optical glass subtract (as seen in fig. \ref{fig:silica_and_glass}). The filter is placed facing the internal side of the box. Transmission and reflection of the dichroic filter is included in the code as an average value between the p- and s-polarization, since it is expected that argon scintillation light is not polarized. The reflection and transmission dependence on the angle of incidence is also taken into account. On top of the filter, a thin layer of PTP is included.
\item {\bf WLS bar:} An acrylic bar of size 93.7~mm x 75.7~mm x 3.5~mm  modelled with the same absorption and emission spectra of EJ-286 from Eljen was positioned as to have its center to coincide with the center of the SiPMs in the height dimension. Its laterals (along the height) are covered with 3M ESR specular reflector.
\item {\bf Silicon Photomultipliers:} 8 SiPMs of size 6~mm x 6~mm were placed in the same positions on the lateral of the X-ARAPUCA frame as the experimental setup. The spectrum of photons which reach the SiPM surface is weighted by the efficiency of the Hamamatsu S13360-6050VE SiPM, as seen in fig.~\ref{SiPMEff}.
\end{itemize}

A few considerations should be made when comparing the results from the simulation and the ones from the experimental setup. For instance, there is no measurement for the quantum yield of PTP in the VUV region available in the literature. The data extends down to the $\sim$200~nm region, where a plateau can be seen~\cite{Lally1996}. This was then extended to the VUV region as a first estimate ($\sim$0.9). Although probably also negligible due to the volume size of the X-ARAPUCA, the simulation does not include an absorption length for liquid argon. Photons traversing WLS bars may be subject to absorption as these materials are reported to have absorption lengths ranging from tenths up to few meters~\cite{Baptista2012, Mufson2013, Howard2018, Mufson2020}. To account for the overall absorption including the effect of the observed micro-cracks reported in section~\ref{sec:analysis} a 0.75~m length was assumed. Recent data published in \cite{xarapuca_milano_tests} has compared the performances of the Eljen WLS bar and one recently developed, FB118. An increase of about 50\% in the number of photons detected by the FB118 with respect to the Eljen one was observed. If we take the FB 118 as the one with a photon yield of about 100\%, the yield of the EJ286 should be adjusted to 0.66, value taken in this simulation. The size of the gap between the SiPM and the bar was also included with a 0.5~mm value. Changing the size of the gap should not be of great impact on the efficiency estimation as long as it is not zero, at least within the considerations made in this simulation~\cite{Paulucci2020}. 

The detection efficiency for the X-ARAPUCA is estimated as $\epsilon_{\text{MC}} = 3.4~\pm~0.6\% $ with an overvoltage of +5.0~V for the SiPMs. The quoted uncertainty is obtained as the square root of the quadrature sum of the statistical fluctuations in the Monte Carlo sampling, PTP quantum yield uncertainty as reported in~\cite{Lally1996}, and variation of WLS bar absorption length value. In the simulation, photon transport processes are not subject to possible small imperfections in the materials. All simulated materials are smooth and without contaminations, unless clearly stated. Therefore, the simulation efficiency estimate for the device is expected to be higher than the values obtained from the experimental setup.
The simulation shows that an increase in efficiency compared to the standard ARAPUCA, that does not include the WLS bar and instead deposits TPB on the dichroic filter facing the inside of the box, is of order of 15-40\%~\cite{Paulucci2020}.

\section{Conclusions}

A complete characterization of the X-ARAPUCA single cell device has been performed with three different ionizing radiations ($\alpha$-particles, $\gamma$-rays and $\mu$) in liquid argon. The device has an acceptance window of 8~cm$~\times$~10~cm and corresponds to about one twenty-fourth of the X-ARAPUCA module which will be installed in the first DUNE far detector module.
%The experimental procedures and instrumentation have been validated through data analysis with a consistent purity for the liquid argon and a precise particle selection over pulse shape discrimination. Together with a complete GEANT4 simulation of the experimental
We measured a detection efficiency ranging from 2.2\% to 2.3\% and from 2.7\% to 3.1\% with  overvoltages of +5.0~V and +5.5~V ($\pm$~0.2~V) respectively. The analysis takes into account the quenching effect of the oxygen and nitrogen contamination of LAr through the measure of the slow scintillation decay time and the cross-talk of the SiPMs. These are not of the same type as the candidates for the first DUNE far detector module, which have a significantly smaller cross-talk probability.

These results are in reasonable agreement with the outcomes of a dedicate and detailed MC simulation of the X-ARAPUCA prototype, if one considers that the simulation is not accounting for non-ideal surfaces of the WLS plate, presence of micro-cracks in the plastic and re-absorptions of the fluor. 

These results are also compatible with the detection efficiency of 2\% reported for the ARAPUCAs in protoDUNE~\cite{first_results_protodune}, if one takes into account the different SiPM coverage (12 SiPMs for the ARAPUCA and 8 SiPMs for the X-ARAPUCA, both 6$\times$6~mm$^2$) and the intrinsic higher collection efficiency of the X-ARAPUCA with respect to the standard ARAPUCA (15\% to 40\% according to MC simulation).

The detection efficiency of 2.2~$\pm$~0.5\%  agrees with that measured for a two-cell X-ARAPUCA prototype with an $^{241}$Am source in 2020~\cite{xarapuca_milano_tests}, where an efficiency of 1.9~$\pm$~0.1\% has been found. The prototype tested in 2020 is essentially the same as the one used in the present work, just with double the size (two dichroic windows).
%The same ratio of the photosensor area to the surface is the same between the two works, because there was 16 SiPMs for the double sized device.
The SiPMs used in the double cell tests are the S14160-6050HS, operated at 2.7 O.V. These SiPMs have a 50\% PDE at this O.V., while the SiPMs used in the present work reach the same QE at +5.0~V O.V..

The tests presented in this work represent an important step in the validation process of the X-ARAPUCA device for its use as the reference photon-detection device for the first DUNE far detector module. The high detection efficiency achieved with the latest versions of the X-ARAPUCA will allow to expand the physics reach of the experiment mainly in the energy region relevant to Supernova neutrinos detection.  

%\appendix
%\section{Some title}
%Please always give a title also for appendices.

\acknowledgments

The authors would like to thank Funda\c c\~ao de Amparo \`a Pesquisa do Estado de S\~ao Paulo (FAPESP) for financial support under grants no 2016/01106-5, 2017/13942-5, and 2019/11557-2. H.~V.~Souza thanks to CNPq/IFGW for the scholarship. V.~L.~Pimentel thanks COLAB, open facilities of CTI Renato Archer and the financial support of the CNPq/PCI program through the M.~C.~Q.~Bazetto fellowship. This work was partially supported by CONACyT research grant: A1-S-23238.

%\paragraph{Note added.} This is also a good position for notes added
%after the paper has been written.


%\begin{thebibliography}{99}

\bibliographystyle{JHEP}
\bibliography{bibliography_henrique}

\providecommand{\href}[2]{#2}\begingroup\raggedright\begin{thebibliography}{10}

\bibitem{propostaARA}
A.~Machado and E.~Segreto, \emph{ARAPUCA a new device for liquid argon
  scintillation light detection}, {\emph{Journal of Instrumentation} {\bfseries
  11} (2016) C02004}.

\bibitem{first_results_protodune}
B.~Abi, A.~A. Abud, R.~Acciarri, M.~Acero, G.~Adamov, M.~Adamowski et~al.,
  \emph{First results on {ProtoDUNE}-{SP} liquid argon time projection chamber
  performance from a beam test at the {CERN} Neutrino Platform},
  \href{http://dx.doi.org/10.1088/1748-0221/15/12/p12004}{\emph{Journal of
  Instrumentation} {\bfseries 15} (dec, 2020) P12004--P12004}.

\bibitem{DUNE_vol4}
B.~Abi, R.~Acciarri, M.~A. Acero, G.~Adamov, D.~Adams, M.~Adinolfi et~al.,
  \emph{Deep Underground Neutrino Experiment (DUNE), Far Detector Technical
  Design Report, Volume IV: Far Detector Single-phase Technology},  2020.

\bibitem{pTP}
T.~DeVol, D.~Wehe and G.~Knoll, \emph{Evaluation of p-terphenyl and
  2,2\textquotesingle\textquotesingle\ dimethyl-p-terphenyl as wavelength
  shifters for barium fluoride},
  \href{http://dx.doi.org/https://doi.org/10.1016/0168-9002(93)90701-I}{\emph{Nuclear
  Instruments and Methods in Physics Research Section A: Accelerators,
  Spectrometers, Detectors and Associated Equipment} {\bfseries 327} (1993) 354
  -- 362}.

\bibitem{eljen_286}
``ELJEN - EJ286.'' \url{https://eljentechnology.com/}.

\bibitem{hmmt_s13360}
``Hamamatsu MPPC S13360.''
  \url{https://www.hamamatsu.com/eu/en/product/type/S13360-6050VE/index.html}.

\bibitem{vikuiti}
``Vikuiti\textsuperscript{\texttrademark} Enhanced Specular Reflector (ESR).''
  \url{https://www.3m.com/}.

\bibitem{alpha_spectrum_tab}
\url{https://www-nds.iaea.org/}.

\bibitem{SiPM_better}
P.~Eckert, H.-C. Schultz-Coulon, W.~Shen, R.~Stamen and A.~Tadday,
  \emph{Characterisation studies of silicon photomultipliers},
  \href{http://dx.doi.org/10.1016/j.nima.2010.03.169}{\emph{Nuclear Instruments
  and Methods in Physics Research Section A: Accelerators, Spectrometers,
  Detectors and Associated Equipment} {\bfseries 620} (Aug, 2010) 217–226}.

\bibitem{Cross_talk_vino}
S.~{Vinogradov}, T.~{Vinogradova}, V.~{Shubin}, D.~{Shushakov} and
  K.~{Sitarsky}, \emph{Probability distribution and noise factor of solid state
  photomultiplier signals with cross-talk and afterpulsing},  in \emph{2009
  IEEE Nuclear Science Symposium Conference Record (NSS/MIC)}, pp.~1496--1500,
  2009.

\bibitem{LAr_fund_properties}
T.~Doke, \emph{{Fundamental Properties of Liquid Argon, Krypton and Xenon as
  Radiation Detector Media}}, .

\bibitem{abudance_dependence}
A.~Hitachi, T.~Takahashi, N.~Funayama, K.~Masuda, J.~Kikuchi and T.~Doke,
  \emph{Effect of ionization density on the time dependence of luminescence
  from liquid argon and xenon},
  \href{http://dx.doi.org/10.1103/PhysRevB.27.5279}{\emph{Phys. Rev. B}
  {\bfseries 27} (May, 1983) 5279--5285}.

\bibitem{model_nuclear_recoil_nl}
D.-M. Mei, Z.-B. Yin, L.~Stonehill and A.~Hime, \emph{A model of nuclear recoil
  scintillation efficiency in noble liquids},
  \href{http://dx.doi.org/10.1016/j.astropartphys.2008.06.001}{\emph{Astroparticle
  Physics} {\bfseries 30} (Aug, 2008) 12–17}.

\bibitem{x_arapuca}
A.~Machado, E.~Segreto, D.~Warner, A.~Fauth, B.~Gelli, R.~M{\'{a}}ximo et~al.,
  \emph{The X-{ARAPUCA}: an improvement of the {ARAPUCA} device},
  \href{http://dx.doi.org/10.1088/1748-0221/13/04/c04026}{\emph{J. Instrum.}
  {\bfseries 13} (apr, 2018) C04026--C04026}.

\bibitem{alpha_equation}
S.~Pommé and B.~C. Marroyo, \emph{Improved peak shape fitting in alpha
  spectra},
  \href{http://dx.doi.org/https://doi.org/10.1016/j.apradiso.2014.11.023}{\emph{Applied
  Radiation and Isotopes} {\bfseries 96} (2015) 148 -- 153}.

\bibitem{cobalt_60}
M.-M. B\'e, V.~Chist\'e, C.~Dulieu, E.~Browne, C.~Baglin, V.~Chechev et~al.,
  \emph{Table of Radionuclides}, vol.~3 of \emph{Monographie BIPM-5}.
\newblock Bureau International des Poids et Mesures, Pavillon de Breteuil,
  F-92310 S\`evres, France, 2006.

\bibitem{pdg}
{\scshape Particle Data Group} collaboration, K.~A. Olive et~al., \emph{{Review
  of Particle Physics}},
  \href{http://dx.doi.org/10.1088/1674-1137/38/9/090001}{\emph{Chin. Phys.}
  {\bfseries C38} (2014) 090001}.

\bibitem{MuonsCecchini2012}
S.~{Cecchini} and M.~{Spurio}, \emph{{Atmospheric muons: experimental
  aspects}},
  \href{http://dx.doi.org/10.5194/gid-2-603-2012}{\emph{Geoscientific
  Instrumentation, Methods and Data Systems Discussions} {\bfseries 2} (Aug.,
  2012) 603--641}, [\href{https://arxiv.org/abs/1208.1171}{{\ttfamily
  1208.1171}}].

\bibitem{nitrogen_contamination_roberto}
R.~Acciarri, M.~Antonello, B.~Baibussinov, M.~Baldo-Ceolin, P.~Benetti,
  F.~Calaprice et~al., \emph{Effects of Nitrogen contamination in liquid
  Argon}, {\emph{Journal of Instrumentation} {\bfseries 5} (2010) P06003}.

\bibitem{Segreto_2021}
E.~Segreto, \emph{Properties of liquid argon scintillation light emission},
  \href{http://dx.doi.org/10.1103/physrevd.103.043001}{\emph{Physical Review D}
  {\bfseries 103} (Feb, 2021) }.

\bibitem{Ettore_tpb}
E.~Segreto, \emph{Evidence of delayed light emission of tetraphenyl-butadiene
  excited by liquid-argon scintillation light},
  \href{http://dx.doi.org/10.1103/PhysRevC.91.035503}{\emph{Phys. Rev. C}
  {\bfseries 91} (Mar, 2015) 035503}.

\bibitem{root_cern}
R.~Brun and F.~Rademakers, \emph{{ROOT: An object oriented data analysis
  framework}},
  \href{http://dx.doi.org/10.1016/S0168-9002(97)00048-X}{\emph{Nucl. Instrum.
  Meth. A} {\bfseries 389} (1997) 81--86}.

\bibitem{LAr_arapuca_test}
E.~Segreto, A.~Machado, L.~Paulucci, F.~Marinho, D.~Galante, S.~Guedes et~al.,
  \emph{Liquid argon test of the ARAPUCA device}, {\emph{Journal of
  Instrumentation} {\bfseries 13} (2018) P08021}.

\bibitem{xarapuca_milano_tests}
C.~Brizzolari, S.~Brovelli, F.~Bruni, P.~Carniti, C.~Cattadori, A.~Falcone
  et~al., \emph{Enhancement of the X-Arapuca photon detection device for the
  {DUNE} experiment},  sep, 2021.
\newblock 10.1088/1748-0221/16/09/p09027.

\bibitem{first_lar_test}
E.~Segreto, A.~Machado, A.~Fauth, R.~Ramos, G.~de~Souza, H.~Souza et~al.,
  \emph{First liquid argon test of the {X}-{ARAPUCA}},
  \href{http://dx.doi.org/10.1088/1748-0221/15/05/c05045}{\emph{Journal of
  Instrumentation} {\bfseries 15} (may, 2020) C05045--C05045}.

\bibitem{Agostinelli2003}
S.~Agostinelli et~al., \emph{Geant4—a simulation toolkit}, {\emph{Nuclear
  Instruments and Methods in Physics Research Section A: Accelerators,
  Spectrometers, Detectors and Associated Equipment} {\bfseries 506} (2003) 250
  -- 303}.

\bibitem{Allison2016}
J.~Allison et~al., \emph{Recent developments in Geant4}, {\emph{Nuclear
  Instruments and Methods in Physics Research Section A: Accelerators,
  Spectrometers, Detectors and Associated Equipment} {\bfseries 835} (2016) 186
  -- 225}.

\bibitem{Lally1996}
C.~Lally, G.~Davies, W.~Jones and N.~Smith, \emph{UV quantum efficiencies of
  organic fluors},
  \href{http://dx.doi.org/https://doi.org/10.1016/0168-583X(96)00318-7}{\emph{Nuclear
  Instruments and Methods in Physics Research Section B: Beam Interactions with
  Materials and Atoms} {\bfseries 117} (1996) 421 -- 427}.

\bibitem{Baptista2012}
B.~Baptista, L.~Bugel, C.~Chiu, J.~Conrad, C.~Ignarra, B.~Jones et~al.,
  ``Benchmarking TPB-coated Light Guides for Liquid Argon TPC Light Detection
  Systems.'' arXiv:1210.3793, 2012.

\bibitem{Mufson2013}
S.~Mufson and B.~Baptista, \emph{Light guide production for {LBNE} and the
  effects of {UV} exposure on {VUV} waveshifter efficiency},
  \href{http://dx.doi.org/10.1088/1748-0221/8/09/c09006}{\emph{Journal of
  Instrumentation} {\bfseries 8} (sep, 2013) C09006--C09006}.

\bibitem{Howard2018}
B.~Howard, S.~Mufson, D.~Whittington, B.~Adams, B.~Baugh, J.~Jordan et~al.,
  \emph{A novel use of light guides and wavelength shifting plates for the
  detection of scintillation photons in large liquid argon detectors},
  \href{http://dx.doi.org/https://doi.org/10.1016/j.nima.2018.06.050}{\emph{Nuclear
  Instruments and Methods in Physics Research Section A: Accelerators,
  Spectrometers, Detectors and Associated Equipment} {\bfseries 907} (2018) 9
  -- 21}.

\bibitem{Mufson2020}
S.~Mufson, B.~Adams, B.~Baugh, B.~Howard, C.~Macias, G.~Cancelo et~al.,
  \emph{Differences in the response of two light guide technologies and two
  readout technologies after an exchange of liquid argon in the dewar},
  \href{http://dx.doi.org/https://doi.org/10.1016/j.nima.2020.164240}{\emph{Nuclear
  Instruments and Methods in Physics Research Section A: Accelerators,
  Spectrometers, Detectors and Associated Equipment} {\bfseries 976} (2020)
  164240}.

\bibitem{Paulucci2020}
L.~Paulucci, F.~Marinho, A.~Machado and E.~Segreto, \emph{A complete simulation
  of the X-{ARAPUCA} device for detection of scintillation photons},
  \href{http://dx.doi.org/10.1088/1748-0221/15/01/c01047}{\emph{Journal of
  Instrumentation} {\bfseries 15} (jan, 2020) C01047--C01047}.

\end{thebibliography}\endgroup

% Please avoid comments such as "For a review'', "For some examples",
% "and references therein" or move them in the text. In general,
% please leave only references in the bibliography and move all
% accessory text in footnotes.

% Also, please have only one work for each \bibitem.

%\end{thebibliography}
\end{document}